\documentclass[10pt,aps,prd,superscriptaddress,nofootinbib,nobibnotes,longbibliography,floatfix,twocolumn]{revtex4-2}

\usepackage{bm}
\usepackage{mathtools,
amsmath,
amssymb,
amsfonts,
mathrsfs,
chngcntr,
multirow}

\let\cc\corresponds
\let\corresponds\relax
\usepackage{mathabx}
\let\corresponds\cc

\usepackage[utf8]{inputenc}
\usepackage[T1]{fontenc}
\usepackage[dvipsnames]{xcolor}
\usepackage[unicode]{hyperref}
\hypersetup{colorlinks=true, citecolor=MidnightBlue,
            linkcolor=MidnightBlue, urlcolor=MidnightBlue, linktocpage=true}
\usepackage[normalem]{ulem}
\usepackage{orcidlink}
\usepackage[capitalize]{cleveref}
\usepackage{float}
\usepackage{comment}

\usepackage{orcidlink}

\newcommand{\orcid}[1]{\href{https://orcid.org/#1}{\includegraphics[width=10pt]{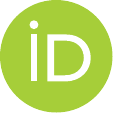}}}

\usepackage{placeins}

\begin{document}

\title{Broken isospectrality in inverted black hole potentials \\ and the emergence of Efimov scaling}

\author{Umer Shahzad \orcid{0009-0007-7483-3063}}
\email{umer.shahzad@student.uni-tuebingen.de}
\affiliation{Theoretical Astrophysics, IAAT, University of T\"ubingen, D-72076 T\"ubingen, Germany}

\author{Sebastian H.\,V\"olkel \orcid{0000-0002-9432-7690}}
\email{sebastian.voelkel@uni-tuebingen.de}
\affiliation{Theoretical Astrophysics, IAAT, University of T\"ubingen, D-72076 T\"ubingen, Germany}

\begin{abstract}
Are the inverted Regge-Wheeler and Zerilli potentials arising in Mashhoon's bound-state approach to black-hole quasinormal modes isospectral?
Extending previous numerical and analytical studies of the inverted Regge-Wheeler potential, we present the first direct computation of the bound-state spectrum of the inverted Zerilli potential.
Although Chandrasekhar proved that the original potentials have identical quasinormal-mode spectra, we show that their inverted counterparts have distinct bound-state spectra.
We also demonstrate that potential inversion destroys their first-order Darboux relation, used by Glampedakis et al.\ to provide a simpler proof of isospectrality.
A continuous interpolation between the potentials reveals substantially larger energy shifts in the interior than at the endpoints, indicating cancellations that suppress the endpoint spectral splitting.
Explaining this behavior requires second-order perturbative contributions, even towards the eikonal limit.
Despite these differences, both spectra exhibit asymptotic geometric scaling characteristic of Efimov physics, first discovered in three-boson systems.
This scaling also applies to other inverted potentials with the same leading tail, including the scalar-field case studied here.
Beyond their theoretical implications, our results raise questions for recently developed methods that compute quasinormal modes with high precision from numerical bound states.
\end{abstract}

\maketitle

\section{Introduction}\label{intr}

Among the surprising findings in black hole perturbation theory is that complicated coupled perturbation equations can be reduced to much simpler wave equations~\cite{Berti:2025hly}.
For the Schwarzschild black hole~\cite{1916SPAW.......189S}, these take the form of two decoupled, time-independent Schrödinger equations for axial and polar perturbations, governed by the Regge-Wheeler~\cite{Regge:1957td} and Zerilli~\cite{Zerilli:1970se} potential barriers, respectively.
Boundary conditions of purely outgoing waves at infinity and purely ingoing waves at the horizon determine a discrete spectrum of quasinormal modes.
Even more surprising is Chandrasekhar's demonstration that the two spectra are identical~\cite{Chandrasekhar:1975zza,Chandrasekhar:1985kt}, later linked to Darboux transformations~\cite{Glampedakis:2017rar}.
This property, known as isospectrality, extends to the Teukolsky equation~\cite{Teukolsky:1973ha} describing perturbations of the Kerr black hole~\cite{Kerr:1963ud}.
Isospectrality is a central property of general relativity that can be violated in alternative theories of gravity.
Detecting such violations in ringdown observations~\cite{Detweiler:1980gk,Dreyer:2003bv,Berti:2005ys} would be a smoking gun for a breakdown of general relativity or the assumptions underlying the Kerr hypothesis~\cite{Israel:1967wq,Carter:1971zc,Hawking:1971vc,Robinson:1975bv}.

Motivated by the close relation to quantum mechanical bound states in potential wells, Mashhoon introduced a pioneering mapping between quasinormal modes of potential barriers and bound-state spectra of the corresponding inverted potentials~\cite{Mashhoon:1982im}.
This mapping was further explored for other potentials in Refs.~\cite{BLOME1984231,Ferrari:1984ozr,Ferrari:1984zz,Churilova:2021nnc}.
The method is widely used in quasinormal mode studies because it provides simple analytic estimates of black hole quasinormal modes.
However, until recently, it was largely limited to a few exactly solvable potentials, as the analytic continuation requires an analytic expression for the bound-state spectrum.

These limitations have been significantly relaxed in recent studies.
Hatsuda~\cite{Hatsuda:2019eoj} proposed a semi-analytic, high-order local WKB expansion around the inverted potential minimum, combined with Borel summation, yielding accurate quasinormal modes for moderate $n$.
A complementary approach, based on numerically constructing a Taylor expansion of the bound-state spectrum in the potential parameters and then analytically continuing this expansion, was proposed in Ref.~\cite{Volkel:2022ewm}.
Its convergence and limitations were explored in Ref.~\cite{Li:2026ptw} and studied for the Schwarzschild-de Sitter black hole in Ref.~\cite{Chen:2026ehv}.
The numerical mapping was further extended using homotopy deformations in Ref.~\cite{Ma:2026bxb} and applied to the Reissner-Nordstr\"om black hole in Ref.~\cite{Zhou:2026kgp}.
However, the mapping does not necessarily recover the complete quasinormal mode spectrum~\cite{Richarte:2024bmm}.

While most works use bound states as a means to compute quasinormal modes, the first explicit presentation of the bound states of the inverted Regge-Wheeler potential was recently reported in Ref.~\cite{Volkel:2025lhe} and confirmed shortly afterwards by analytic results for two limiting cases~\cite{Hod:2025scz}.
The physical interpretation of the quasinormal mode spectrum, even for Schwarzschild black holes, has attracted substantial interest~\cite{Bekenstein:1974jk,Mukhanov:1986me,Hod:1998vk,Maggiore:2007nq}, and recent advances in the bound-state mapping might offer unexpected insights~\cite{Hatsuda:2019eoj,Volkel:2022ewm,Richarte:2024bmm,Volkel:2025lhe,Hod:2025scz,Li:2026ptw,Chen:2026ehv,Ma:2026bxb,Zhou:2026kgp}. 

Our study presents the first direct computation of the bound states of the inverted Zerilli potential.
It reveals the arguably counterintuitive result that its spectrum differs from that of the inverted Regge-Wheeler potential.
Thus, inverting the two isospectral potential barriers yields non-isospectral wells, raising the question of how quasinormal-mode isospectrality relates to the bound-state spectra under analytic continuation.

We further demonstrate that the conditions for the first-order Darboux transformation between the two isospectral potential barriers, considered in Ref.~\cite{Glampedakis:2017rar} in the context of quasinormal modes, do not carry over to the corresponding inverted potentials.
Although this does not formally prove non-isospectrality, it shows that any isospectrality cannot arise from a first-order Darboux relation. 
Although the energy differences are very small, they are not well approximated by first-order perturbation theory.
Surprisingly, a continuous interpolation between the two potentials reveals much larger energy shifts for most bound states in an intermediate regime than at the endpoint.

Despite these differences, we identify large-$n$ asymptotic properties of the bound states of both potentials that resemble the geometric scaling known in Efimov physics~\cite{Efimov:1970zz,Naidon:2016dpf}.
Efimov discovered that three identical bosons with resonant short-range interactions can support an infinite tower of three-body bound states, even when no two-body bound state exists.
This behavior is governed by an effective attractive inverse-square potential.
Experimental evidence for Efimov states has since been obtained in ultracold cesium gases~\cite{Kraemer:2006pnv} and helium trimers~\cite{Kunitski:2015qth}; see Ref.~\cite{Naidon:2016dpf} for a review.
Efimov-type scaling has also been discussed in the distinct setting of holographic models involving charged anti-de Sitter black holes~\cite{Iqbal:2010eh}.
In our case, the common inverse-square tail determines the asymptotic spacing, while the details of each potential determine the spectral offset.

\section{Methods}\label{meth}

\subsection{Inverted black hole potentials}

In Mashhoon's approach~\cite{Mashhoon:1982im}, the squared QNM frequencies $\omega_n^2$ of a potential barrier $V(x)$ in the time-independent Schr\"odinger equation
\begin{align}\label{schrodinger}
\frac{\text{d}^2}{\text{d}x^2}\psi (x) + \left[\omega_n^2 - V(x) \right] \psi(x) = 0\,,
\end{align}
are related to the bound-state eigenvalues $E_n$ of the inverted potential $-V(x)$.
Applying the complex coordinate transformation $x\rightarrow -ix$ to Eq.~\eqref{schrodinger} and requiring
\begin{align}
V(x,P)=V(-ix,P')\,,
\end{align}
yields a mapping between the potential parameters, $P'=\pi(P)$.
Defining $\Omega_n^2(P)\equiv -E_n(P)$ for the bound-state problem, the QNM frequencies follow from
\begin{align}
\label{bound_to_qnm}
\omega_n(P)=\Omega_n\!\left(\pi^{-1}(P)\right)\,.
\end{align}
Here, $P$ denotes the parameters of the original potential, while $\pi$ depends on its specific form.
In the following, we use only the formal correspondence between the QNM and bound-state spectra, without constructing this mapping explicitly.

For axial and polar perturbations of the Schwarzschild black hole, Eq.~\eqref{schrodinger} is expressed in the tortoise coordinate $x(r) = r + 2\,M \ln\left[r/(2M)-1\right]$, where $M$ is the black hole mass.
This maps the exterior region $r \in (2M, \infty)$ to $x \in (-\infty, \infty)$.
The Regge-Wheeler and Zerilli potentials are given by
\begin{align}\label{potential_RW}
V^{\text{RW}}_{\ell}(r) = f(r) \left(\frac{\ell(\ell+1)}{r^2}-\frac{6M}{r^3} \right)\,,
\end{align}
and
\begin{align}\label{potential_Z}
V^{\text{Z}}_{\ell}(r) = f(r)\, \frac{2\lambda^2(\lambda+1)\,r^3+6\lambda^2 Mr^2+18\lambda M^2 r+18M^3}{r^3(\lambda r+3M)^2}\,,
\end{align}
with
\begin{align}
    f(r) = \left(1- \frac{2M}{r}\right), \qquad
    \lambda = \frac{(\ell -1)(\ell +2)}{2}\,,
\end{align}
where $\ell(\ell+1)$ is the angular separation constant.
The effective potential of a test scalar field is given by
\begin{align}\label{potential_RW}
V^{\text{scalar}}_{\ell}(r) = f(r) \left(\frac{\ell(\ell+1)}{r^2}+\frac{2M}{r^3} \right)\,.
\end{align}
At large $x$, all potentials share the inverse-square tail $V_\ell(x)\sim \ell(\ell+1)/x^2$, with subleading logarithmic corrections from the tortoise-coordinate transformation.
Bound states of the inverted potentials correspond to solutions that decay towards $x \rightarrow \pm \infty$.

To further explore the spectral properties of the two potentials in Sec.~\ref{app_spectral_splitting}, we introduce the linear interpolation
\begin{equation}
\label{interpolative_potential}
V_{\ell}^{\rm{s}}(r,\epsilon)
= V_{\ell}^{\rm{s}}(r)
+ \epsilon\,\Delta V_{\ell}^{\rm{s}}(r),
\qquad \epsilon\in[0,1],
\end{equation}
where $\rm{s}\in\{\rm{RW},\rm{Z}\}$ labels the reference potential and
\begin{align}
\Delta V_{\ell}^{\rm{RW}}(r)
&= V_{\ell}^{\rm{Z}}(r) - V_{\ell}^{\rm{RW}}(r),
\\
\Delta V_{\ell}^{\rm{Z}}(r)
&=  V_{\ell}^{\rm{RW}}(r) - V_{\ell}^{\rm{Z}}(r).
\end{align}
Thus, $\epsilon=0$ recovers the reference potential, while $\epsilon=1$ yields the potential of the other sector.
We write the bound-state spectrum of the corresponding inverted potential as
\begin{align}\label{method_en_shift}
 E^{\rm s}_{\ell n}(\epsilon) = E^{\rm s}_{\ell n} \left[ 1 +  \delta E^{\rm s}_{\ell n}(\epsilon)\right],
\end{align}
where $\delta E^{\rm s}_{\ell n}(\epsilon)$ is the relative energy shift from the reference spectrum at $\epsilon=0$. 
To characterize the energy shifts at small $\epsilon$, we approximate
the eigenvalues by a third-order Taylor expansion about $\epsilon=0$,
\begin{align}\label{method_en_shift2}
 E^{\rm s,Taylor}_{\ell n}(\epsilon)
 &\equiv \sum_{k=0}^{3} \frac{\epsilon^k}{k!}\,d_k\,,
 &
 d_k &\equiv
 \left.
 \frac{\mathrm{d}^k E^{\rm s}_{\ell n}}{\mathrm{d}\epsilon^k}
 \right|_{\epsilon=0}\,.
\end{align}

\subsection{Breaking of first-order Darboux isospectrality}

The first-order Darboux transformation maps solutions of Eq.~\eqref{schrodinger} to those of a partner potential and can establish isospectrality under suitable conditions.
It was used in Ref.~\cite{Glampedakis:2017rar} to give a simple derivation of Regge-Wheeler-Zerilli isospectrality, previously established by Chandrasekhar~\cite{Chandrasekhar:1975zza,Chandrasekhar:1985kt}.
Below, we show that this transformation cannot establish isospectrality between the corresponding inverted potentials.
This alone does not prove that isospectrality is broken, as more general classes of isospectral potentials exist.

The first-order Darboux transformation between functions $Y(x)$ and $y(x)$,
\begin{align}
Y = y^\prime + f y,
\end{align}
can help establish isospectrality of the potentials in
\begin{align}
Y^{\prime \prime} + Q Y = 0, \qquad y^{\prime \prime} + q y = 0,
\end{align}
provided the constraints
\begin{align}\label{Darboux_constraints}
f = \frac{(Q+q)^\prime}{2(Q-q)}, \qquad f^\prime = \frac{1}{2}(q-Q),
\end{align}
are satisfied~\cite{Glampedakis:2017rar}.
Identifying $q=\lambda-V_\ell^{\rm RW}(r)$ and $Q=\lambda-V_\ell^{\rm Z}(r)$ gives the original potential-barrier problem, for which the existence and explicit form of $f$ were established in Ref.~\cite{Glampedakis:2017rar}.

Next, we define $\tilde{q}=E+V_\ell^{\rm RW}(r)$ and $\tilde{Q}=E+V_\ell^{\rm Z}(r)$ for the corresponding inverted-potential equations, with spectral parameter $E$, solutions $\tilde{y}$ and $\tilde{Y}$, and a hypothetical first-order Darboux generator $\tilde{f}$.
The corresponding constraints are
\begin{align}\label{Darboux_constraints_2}
\tilde{f} = \frac{(\tilde{Q}+\tilde{q})^\prime}{2(\tilde{Q}-\tilde{q})}, \qquad \tilde{f}^\prime = \frac{1}{2}(\tilde{q}-\tilde{Q}).
\end{align}
The constraints in Eqs.~\eqref{Darboux_constraints} and~\eqref{Darboux_constraints_2} are independent of the spectral parameters $\lambda$ and $E$, which cancel in the differences and vanish under differentiation.

Inverting the potentials leaves the first constraint unchanged but reverses the sign of the second.
Thus, $f = \tilde{f}$ and $f^\prime = -\tilde{f}^\prime$, which together require the two potentials to be identical.
Since the Regge-Wheeler and Zerilli potentials differ and their Darboux relation is established~\cite{Glampedakis:2017rar}, their inverted counterparts cannot be first-order Darboux partners.
More generally, this argument applies to any two distinct potential barriers related by a first-order Darboux transformation.

\subsection{Numerical methods for bound states}\label{numerial_method_sec}

We computed the bound states of the inverted potentials primarily using a shooting method, with independent validation by a pseudospectral method.
Both methods are briefly described below, with further details in Appendix~\ref{app_numerics}.

For the shooting method, we start with an energy guess and numerically integrate Eq.~\eqref{schrodinger} from two boundaries toward a common matching point, imposing bound-state boundary conditions.
This yields two solutions, $\psi_\pm$, whose Wronskian is evaluated at the matching point as $W(\psi_+,\psi_-)= \psi_-\psi_+^{'} - \psi_+ \psi_-^{'} $. At an eigenvalue, the solutions are linearly dependent and their Wronskian vanishes.
The roots of the Wronskian therefore determine the eigenvalue spectrum.

For the pseudospectral method, we constructed a matrix representation of the Hamiltonian $\hat{H} = -{\rm d^2}/{{\rm d}x^2} + V(x)$ using a Chebyshev differentiation matrix on a Chebyshev-Gauss-Lobatto (CGL) grid.
The bound-state energies correspond to eigenvalues of this matrix.
For further details on pseudospectral methods, see Ref.~\cite{trefethen2000spectral}.

\section{Application and results}\label{app}

\subsection{Bound states of inverted potentials}

\begin{figure}[t]
    \centering
    \includegraphics[width=1.0\linewidth]{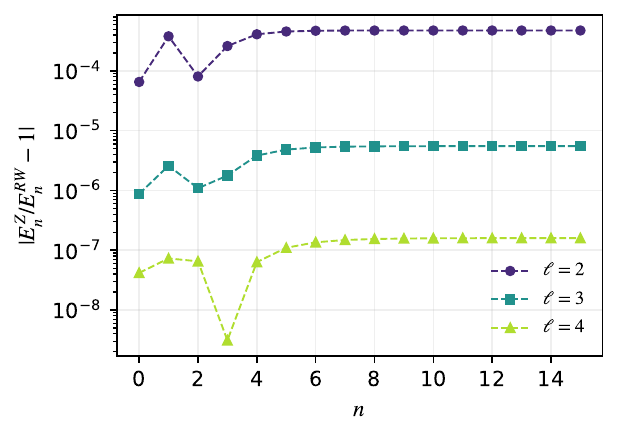}
    \caption{Relative difference between bound states of inverted Regge-Wheeler and Zerilli potential for $\ell = 2 , 3 \, \text{and}\,\, 4$.}
\label{fig:regge_zerilli_relative_difference}
\end{figure}

Figure~\ref{fig:regge_zerilli_relative_difference} shows the relative differences between the bound-state energies of the inverted Regge-Wheeler and Zerilli potentials for different $\ell$ and $n$. Numerical values, including those for the scalar potential, are listed in Table~\ref{tab:master-bound-state-data}. Selected eigenfunctions and additional results are presented in Appendix~\ref{app_add_results}.

Increasing $\ell$ deepens the potential wells and shifts states with fixed $n$ to more negative energies, whereas increasing $n$ at fixed $\ell$ produces progressively more weakly bound states. Low-lying states are concentrated near the potential minimum. In the eikonal regime with $n\ll\ell$, their energies admit a harmonic approximation with anharmonic corrections.

At large $n$, the eigenfunctions extend farther into the asymptotic region, where the long-range tail of the potential governs the spectrum. The energies approach zero exponentially with increasing $n$ at fixed $\ell$. For the Regge-Wheeler and scalar potentials, this behavior is supported by excellent agreement between the numerical results and Hod's asymptotic prediction~\cite{Hod:2025scz}.

Our central finding is that the inverted Regge-Wheeler and Zerilli potentials are not isospectral. For $\ell=2$, the relative energy difference is approximately $6.5\times10^{-5}$ at $n=0$ and approaches $4.7\times10^{-4}$ at large $n$. Over the multipole range investigated, these differences decrease by roughly one to two orders of magnitude with each increment in $\ell$, qualitatively consistent with the common leading-order form of the potentials in the eikonal limit. The numerical results suggest a finite, nonzero relative splitting as $n\to\infty$ at fixed $\ell$, but a vanishing relative splitting as $\ell\to\infty$ at fixed $n$. We examine the large-$n$ limit in Sec.~\ref{app_efimov}.

To establish that the small spectral differences are numerically resolved, we present convergence tests and comparisons between the two independent numerical methods in Appendix~\ref{app_numerics}, together with analytical checks where available. These tests show that the reported differences exceed the estimated numerical uncertainties by several orders of magnitude. Independent numerical validation is particularly important for the Zerilli potential, which is not covered by Hod's analytical results.

\subsection{Properties of the spectral splitting}\label{app_spectral_splitting}

\begin{figure}[t]
    \centering
    \includegraphics[width=1.0  \linewidth]{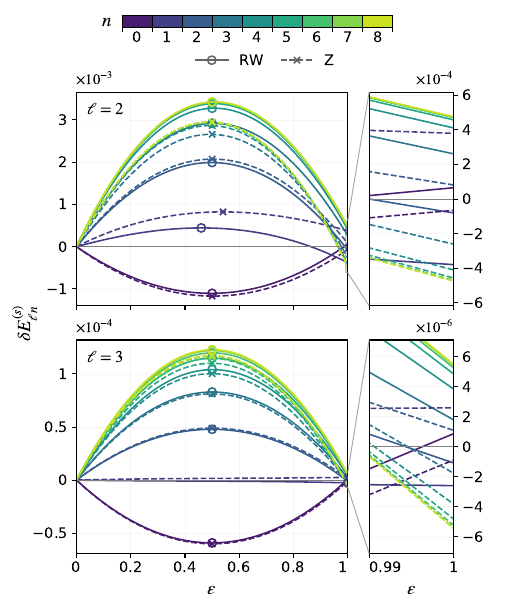}
    \caption{Relative energy shift $\delta E^{\mathrm{s}}_{\ell n}(\epsilon)$, defined in Eq.~\eqref{method_en_shift}, for $\ell=2,3$ (top to bottom) and $n=0,\ldots,8$ (colors). Solid and dashed curves correspond to $\mathrm{s}=\mathrm{RW}$ and $\mathrm{s}=\mathrm{Z}$, respectively. Circles (RW) and crosses (Z) mark the largest absolute relative shift among the sampled points on each curve. The right panels show the near-endpoint behavior for $0.99\leq\epsilon\leq1$.}
    \label{fig:epsilon_dependent_change}
\end{figure}

To assess whether linear perturbation theory captures the spectral shifts, we examine the relative energy shifts $\delta E^{\rm s}_{\ell n}(\epsilon)$ along the interpolations $V_{\ell}^{\rm s}(r,\epsilon)$. Figure~\ref{fig:epsilon_dependent_change} shows the results for $\ell=2,3$, computed using the same numerical methods. For clarity, we omit larger multipoles from the figure and describe their behavior below. By construction, both relative energy shifts vanish at $\epsilon=0$. Most curves are approximately parabolic, with the largest absolute shifts near $\epsilon=1/2$ and nearly linear behavior close to the endpoints. As $\epsilon\to1$, the shifts return to small, generally nonzero values, resolved in the enlarged panels.

For a given $\ell$, the shifts exhibit a clear ordering with $n$. The lowest modes exhibit negative shifts over most of the interpolation range, whereas higher modes generally exhibit positive shifts. The curves increase systematically with $n$ and appear to approach a limiting profile. This ordering is most evident away from the endpoint, where small differences between the Regge-Wheeler and Zerilli results remain visible. Remarkably, the approximately parabolic profile persists as $\ell$ increases, even as its amplitude decreases. The endpoint shifts become progressively smaller relative to the largest shifts in the interior.

In the eikonal regime, modes with $n\ll\ell$ should be well described by a harmonic approximation around the potential minimum, allowing analytical estimates of their energy shifts. However, this local approximation does not directly explain the apparent convergence of the energy-shift curves at large $n$ and fixed $\ell$. The suppression of the endpoint shifts and the persistence of the approximately parabolic profile beyond the regime of validity of the harmonic approximation therefore deserve further analytical investigation.

Finally, the Taylor coefficients $d_k$ support the approximately parabolic dependence of $E^{\rm s}_{\ell n}(\epsilon)$ described by Eq.~\eqref{method_en_shift2}. For large $\ell$, we find numerically that $d_1 \approx -d_2/2$ and
\begin{align}
 \left|\frac{d_3}{6}\right|
 \ll
 \left|d_1\right| \approx \left|\frac{d_2}{2}\right|\,.
\end{align}
The linear and quadratic contributions therefore nearly cancel at $\epsilon=1$, while the cubic contribution remains small compared with either lower-order term. To second order, the relative energy shift takes the form $\delta E^{\rm s}_{\ell n}(\epsilon) \approx d_1 \epsilon(1-\epsilon)/E^{\rm s}_{\ell n}(0)$. These results support a second-order description of the energy curves and show that linear perturbation theory alone is insufficient. Higher-order terms may nevertheless remain important for the small residual endpoint splitting, especially at small $\ell$. Further details are shown in Fig.~\ref{fig:summary_parabola_fit_coefficients} in Appendix~\ref{app_add_results}.

\subsection{Asymptotic Efimov spectra}\label{app_efimov}

Next we explore the large-$n$ asymptotic behavior of the bound states of the Regge-Wheeler, Zerilli, and scalar potentials at fixed $\ell$, and its connection to Efimov physics. For each potential $\rm s$, we write the asymptotic spectrum as
\begin{align}
\label{effimov_spectrum}
E_{\ell n}^{\rm s}
\sim
-E_*
\exp\!\left[
-\frac{2\pi}{s^{\infty}_{\ell}}
\left(n+\delta^{\rm s}_{\ell}\right)
\right],
\qquad n\to\infty .
\end{align}
Here, $s^{\infty}_{\ell}=\sqrt{\ell(\ell+1)-1/4}$ determines the common asymptotic logarithmic level spacing, while $\delta^{\rm s}_{\ell}$ specifies the potential-dependent spectral offset. We fix the common reference energy scale to $E_*=1/M^2$ in geometric units. With this convention, the asymptotic spectrum is characterized by only two parameters, the spacing parameter $s^\infty_\ell$ and the offset $\delta^{\rm s}_{\ell}$. Rearranging Eq.~\eqref{effimov_spectrum}, we get
\begin{equation}
    \label{potential_dependent_phase}
    \delta^{\rm s}_{\ell n} = - \left(n + \frac{s^\infty_\ell }{2 \pi} \ln \left| \frac{E^{\rm s}_{\ell n}}{E_*} \right| \right)\,.
\end{equation}

The potential-dependent phase $\delta^{\rm s}_{\ell}$ characterizes the relative offset of the different Efimov spectra on a logarithmic energy scale. For a given potential and $\ell$, it can be extracted from the large-$n$ numerical spectrum using Eq.~\eqref{potential_dependent_phase} or, where available, from Hod's analytical formula [Eq.~(36) of Ref.~\cite{Hod:2025scz}], which covers the scalar, vector, and Regge-Wheeler potentials, but not Zerilli. To monitor convergence to the asymptotic Efimov spacing, we introduce
\begin{align}\label{effimov_spacing}
s^{\rm s}_{\ell n}
\equiv
\frac{2\pi}
{\ln\!\left[E_{\ell n}^{\rm s}/E_{\ell,n+1}^{\rm s}\right]}\,.
\end{align}

\begin{figure}[t]
    \centering
    \includegraphics[width=1.0\linewidth]{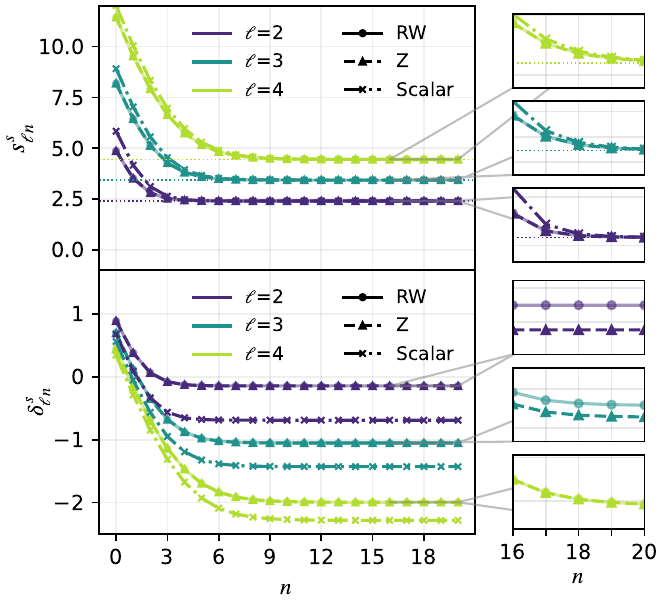}
    \caption{Geometric scaling factors $s^{\rm s}_{\ell n}$ Eq.~\eqref{effimov_spacing} (top panel) and potential dependent phase $\delta^s_{\ell n}$ Eq.~\eqref{potential_dependent_phase} (bottom panel) for the inverted Regge-Wheeler, Zerilli, and scalar potentials with $\ell=2,3,4$. Colors distinguish the multipoles, while circles, triangles, and crosses denote the Regge-Wheeler, Zerilli, and scalar potentials, respectively. The dotted horizontal lines in top panel indicate the asymptotic values $s^\infty_{\ell}$. }
    \label{fig:efimov_spacing_constant}
\end{figure}

Figure~\ref{fig:efimov_spacing_constant} shows $s^{\rm s}_{\ell n}$, $s^\infty_\ell$, and $\delta^{\rm s}_{\ell n}$ for all three potentials and different $\ell$. In each case, $s^{\rm s}_{\ell n}$ rapidly approaches $s^\infty_\ell$, with slower convergence as $\ell$ increases, consistent with the more rapid delocalization of eigenfunctions at smaller $\ell$. The phases also rapidly approach constant values. The Regge-Wheeler and Zerilli phases nearly coincide and are clearly distinct from the scalar phase, with smaller differences between the potentials as $\ell$ increases.

The ratio of the Zerilli and Regge-Wheeler bound-state energies therefore approaches an $n$-independent limit,
\begin{align}
\lim_{n\to\infty}\frac{E_{\ell n}^{\rm Z}}{E_{\ell n}^{\rm RW}}
=
\exp\left[
-\frac{2\pi}{s^\infty_{\ell}}
\left(\delta^{\rm Z}_{\ell}-\delta^{\rm RW}_{\ell}\right)
\right].
\end{align}
The phase difference remains nonzero within our numerical accuracy for the multipoles investigated; see Table~\ref{tab:master-bound-state-data}. Thus, the common asymptotic spacing does not imply a vanishing relative spectral splitting.

\section{Conclusions}\label{concl}

We have presented the first direct computation of the bound states of the inverted Zerilli potential and shown that its spectrum differs from that of the inverted Regge-Wheeler potential.
The latter was studied numerically in Ref.~\cite{Volkel:2025lhe}, with analytical results in limiting regimes subsequently obtained by Hod~\cite{Hod:2025scz}.
Neither study explicitly treated the inverted Zerilli potential.
Extending the analytical treatment to this case would provide an important independent validation of the spectral splitting found here.

The loss of isospectrality is consistent with our finding that potential inversion destroys the first-order Darboux relation underlying the isospectrality of the original barriers.
The absence of this relation alone, however, does not imply distinct spectra.
Our results show that identical quasinormal-mode spectra do not entail identical bound-state spectra for the inverted potentials.

The spectral splitting, including its relative suppression towards the eikonal limit, reveals further structure.
Although the energy-level shifts are generally small, first-order perturbation theory is insufficient to describe them even when $\ell$ grows at fixed small $n$.
A linear interpolation between the two potentials reveals that the largest absolute energy shift relative to the Regge-Wheeler endpoint typically occurs near the midpoint.

At large $n$ and fixed $\ell$, the bound-state spectrum exhibits discrete scale invariance familiar from Efimov physics.
At leading order, the energy levels scale geometrically: the logarithmic spacing of consecutive binding energies approaches a constant set by the coefficient of the leading attractive inverse-square tail.
This constant is the same for the Regge-Wheeler and Zerilli potentials, and the scaling extends to other cases, as we demonstrate for test scalar fields.
The same asymptotic structure can arise in other nonrotating, asymptotically flat spacetimes whose effective potentials share this leading tail and differ only at subleading orders in the $1/r$ expansion.
The overall energy scale, however, depends on the potential details and differs between the cases considered here.
Thus, a common asymptotic spacing does not imply vanishing relative spectral splitting.
In Efimov physics, the corresponding spectral offset encodes the properties of short-range interactions. 
It remains an open question whether this shared spectral structure points to a deeper connection with Efimov systems.

Schwarzschild perturbations can also be described by the Bardeen-Press equation~\cite{Bardeen:1973xb}, which coincides with the Teukolsky equation~\cite{Teukolsky:1973ha} in the nonrotating limit.
The Bardeen-Press potential is complex valued, frequency dependent, and long ranged, unlike the real, frequency-independent, short-range Regge-Wheeler and Zerilli potentials.
Its relation to these potentials requires a generalized rather than an ordinary first-order Darboux transformation~\cite{Glampedakis:2017rar}.
Understanding how such transformations interact with potential inversion and analytic continuation may help extend the inverted-potential approach to rotating black holes beyond the slow-rotation and eikonal approximations~\cite{Ferrari:1984ozr,Ferrari:1984zz}.

Recent developments and critical assessments of methods for computing quasinormal modes from bound states~\cite{Hatsuda:2019eoj,Volkel:2022ewm,Richarte:2024bmm,Li:2026ptw,Chen:2026ehv,Ma:2026bxb,Zhou:2026kgp} lend renewed importance to the questions raised by our findings.
Under what conditions does analytic continuation recover identical quasinormal-mode spectra from the distinct bound-state spectra of inverted isospectral barriers?
Can the choice among isospectral barrier representations improve the convergence and accuracy of numerical continuation methods?

\acknowledgments

The authors thank Kostas D.\,Kokkotas for useful discussions. 
The authors acknowledge the use of OpenAI’s GPT-6 Astra to provide feedback on the clarity, structure, and presentation of the manuscript and to assist in checking mathematical reasoning based on ideas, references, and numerical methods supplied by the authors. The authors reviewed the suggested revisions and independently verified the mathematical arguments against the original references and their own calculations. The authors take full responsibility for the manuscript, including all analytical and numerical calculations.

\bibliography{literature}

\appendix

\section{Details for numerical methods}\label{app_numerics}

Here we provide additional technical details on both numerical methods and present convergence tests.

\subsection{Shooting method}

The shooting method has two key parameters controlling accuracy: the boundary placement and the ODE solver's relative tolerance.
Tighter tolerances generally improve accuracy until machine precision becomes limiting.
Boundary placement, however, requires more care.

For a bound state with energy $E_n < 0$, the wavefunction decays exponentially in the classically forbidden region as $\psi(x) \sim \exp(-\sqrt{|E_n|}\,|x|)$ for $|x| \gg |x_t|$, where $x_t$ is the classical turning point.
Placing the boundary too far from $x_t$ can cause the exponentially growing solution to dominate numerically, leading to overflow and instability.
Placing it too close introduces errors because the asymptotic approximation is not yet valid.

\begin{figure}
    \centering
    \includegraphics[width=1.0\linewidth]{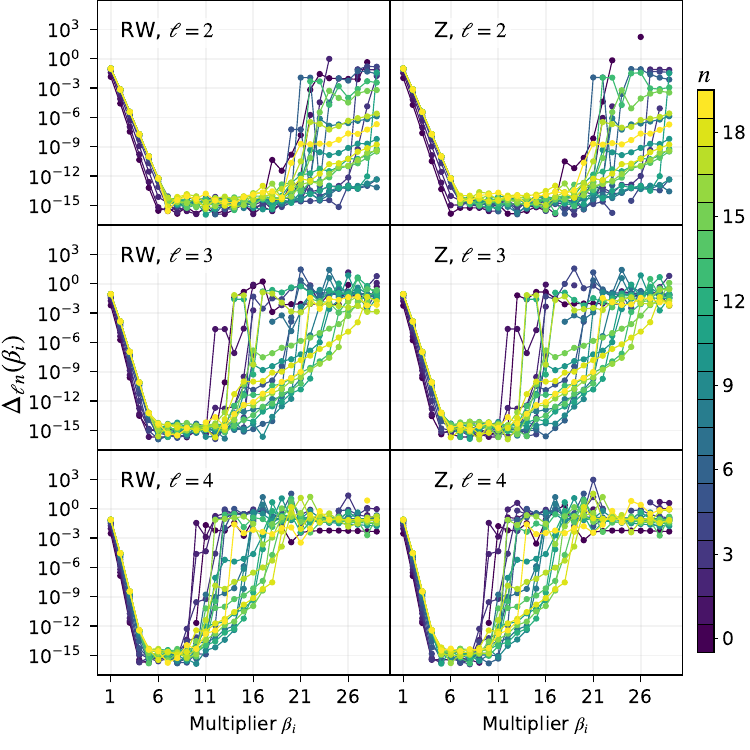}
    \caption{Adjacent-multiplier relative energy differences,
    $\Delta_{\ell n}(\beta_i)=|E_{\ell n}(\beta_{i+1})-E_{\ell n}(\beta_i)|/
    |E_{\ell n}(\beta_{i+1})|$, for the Regge-Wheeler (left) and Zerilli (right)
    potentials. Rows correspond to $\ell=2,3,4$; color indicates energy level $n$.}
    \label{fig:multiplier_convergence}
\end{figure}

We therefore determine the boundary location adaptively for each trial energy.
Using the large-$r$ asymptotic form $V(r) \sim \ell(\ell+1)/r^2$, we estimate the classical turning point $x_t(E_n) = \sqrt{\ell(\ell+1)/|E_n|}$ and place the boundaries at $x_{\pm} = \pm \beta\, x_t$, where $\beta$ is a dimensionless multiplier.
We tested $1 \leq \beta \le 30$ and found that the optimal value of $\beta$ depends weakly on $\ell$: smaller values introduce boundary errors, while larger values lead to numerical overflow. Based on this scan, we adopt $\beta = 8$ for $\ell = 2$ and $\beta = 5$ for $\ell = 3$ and $\ell = 4$, which we found to give the best accuracy for each respective case.
We validated this choice against Hod's analytical expression~\cite{Hod:2025scz}, the results of Ref.~\cite{Volkel:2025lhe}, and our convergence tests in Figure~\ref{fig:multiplier_convergence}.

We integrated the Schr\"odinger equation~\eqref{schrodinger} using an adaptive fifth-order Runge-Kutta method (Dormand-Prince)~\cite{dormand1980}.
The solutions integrated from $x_+$ and $x_-$ were matched at $x = 0$, where we evaluated the Wronskian $W(\psi_+,\psi_-)$.
Eigenvalues were found using the Newton-Raphson method to locate zeros of the Wronskian, with its energy derivative computed numerically using second-order central finite differences.

\begin{figure}[t]
    \centering
    \includegraphics[width=1.0\linewidth]{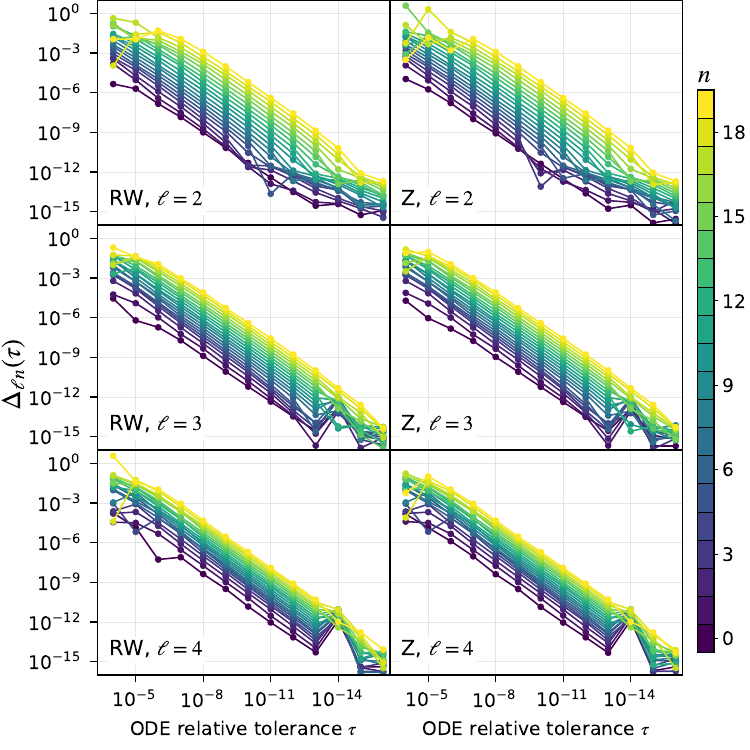}
    \caption{Self-convergence of shooting-method eigenvalues for the inverted Regge–Wheeler and Zerilli potentials , left and right panels, respectively. Panels show $\ell=2,3,4$ from top to bottom; colors indicate the energy level $n\leq20$. Relative differences $\Delta_{\ell n} (\tau) =|E_{\ell n}(\tau)-E_{\ell n}(10^{-17})|/|E_{\ell n}(10^{-17})|$ are plotted against the ODE relative tolerance $\tau$. The reference uses the tightest saved tolerance, $10^{-17}$}
    \label{summary_ode_self_convergence_l234_two_column}
\end{figure}

We tested the self-convergence of the shooting method by tightening the ODE relative tolerance and comparing each eigenvalue with its value at $\mathrm{rtol}=10^{-17}$.
Figure~\ref{summary_ode_self_convergence_l234_two_column} shows the results for the inverted Regge--Wheeler and Zerilli potentials.
The relative differences generally decrease as the tolerance tightens, eventually reaching a numerical floor for many low-lying levels, indicating numerical stability.

\subsection{Pseudospectral method}

As an independent check on the shooting method, we also compute bound states using a pseudospectral (collocation) method.
Such methods have been widely used for eigenvalue problems since the early work of Orszag~\cite{orszag1972} and were popularized by Trefethen's \emph{Spectral Methods in MATLAB}~\cite{trefethen2000spectral}.
Unlike spectral methods formulated in coefficient space, pseudospectral methods work directly in physical space, simplifying the construction of operator matrices for variable potentials while retaining exponential convergence for sufficiently regular solutions.

We discretize the Regge-Wheeler equation on Chebyshev-Gauss-Lobatto grid points $\{\xi_i\}$ in the computational coordinate $\xi \in [-1,1]$, related to the physical coordinate $x$ by
\begin{equation}
x(\xi) = a \frac{\sinh(\gamma \xi)}{\sinh(\gamma)}.
\end{equation}
The derivative relations are
\begin{equation}
\frac{d\xi}{dx} = \frac{\sinh(\gamma)}{\gamma a \cosh(\gamma \xi)}, \qquad
\frac{d^2\xi}{dx^2} = -\frac{\sinh^2(\gamma) \sinh(\gamma \xi)}{\gamma a^2 \cosh^3(\gamma \xi)}.
\end{equation}
Using the chain rule, the second derivative in $x$ is expressed in terms of derivatives in $\xi$ as
\begin{equation}
\frac{d^2}{dx^2} = \left(\frac{d\xi}{dx}\right)^2 \frac{d^2}{d\xi^2} + \frac{d^2\xi}{dx^2} \frac{d}{d\xi}.
\end{equation}

The Chebyshev differentiation matrices $D_\xi$ and $D_{\xi^2}$ are constructed on the Gauss-Lobatto grid (see, e.g., Chapter 6 of Ref.~\cite{trefethen2000spectral}), and the operator matrix for the Hamiltonian is built as
\begin{equation}
L = -g_1^2 D_{\xi^2} - g_2 D_\xi + V(x(\xi)),
\end{equation}
where $g_1 = d\xi/dx$ and $g_2 = d^2\xi/dx^2$ are the metric factors from the coordinate transformation. Bound states are obtained as eigenvalues of this matrix. Since the wavefunction decays exponentially at large distances, we impose Dirichlet boundary conditions by excluding the boundary grid points, i.e., we diagonalize the interior submatrix $L[1:-1, 1:-1]$.

\begin{figure}
    \centering
    \includegraphics[width=1.0\linewidth]{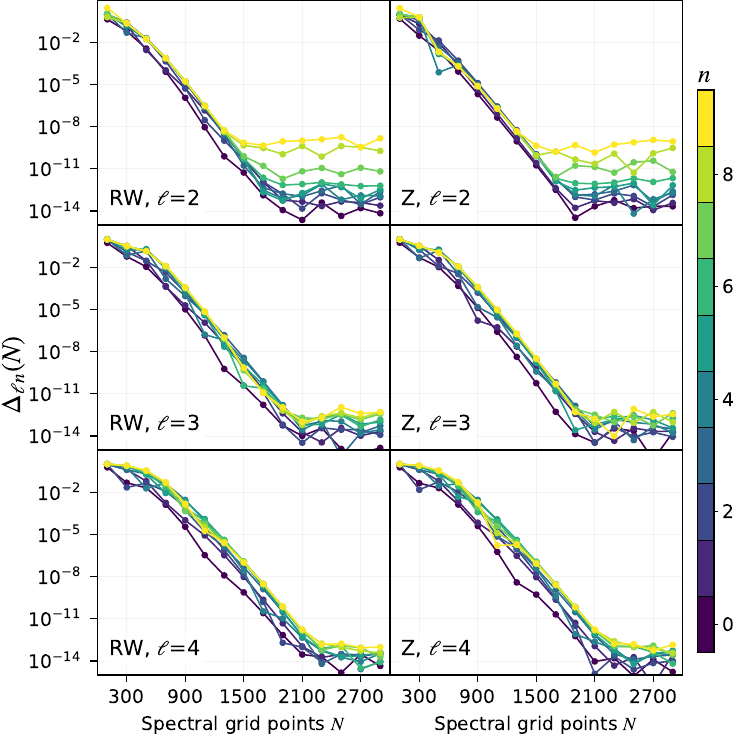}
    \caption{Spectral-grid self-convergence of the inverted Regge-Wheeler (left) and Zerilli (right) eigenvalues. Rows show $\ell=2,3,4$ from top to bottom, and color identifies energy level $n$. The plotted relative difference is $\Delta_{\ell n}(N)=|E_{\ell n}(N)-E_{\ell n}(3100)|/|E_{\ell n}(3100)|$, where $N$ is the number of spectral grid points and $N=3100$ is the finest-grid reference.}
    \label{summary_spectral_self_convergence_l234_two_column}
\end{figure}

The method depends on three numerical parameters: the number of grid points $N$, the boundary location $a$, and the grid-stretching parameter $\gamma$.
The transformation maps $\xi \in [-1,1]$ to $x \in [-a,a]$, while $\gamma$ controls the grid-point distribution.
In the limit $\gamma = 0$, the mapping reduces to $x = a\xi$.
Increasing $\gamma$ concentrates more points near the center ($x=0$) and fewer near the boundaries, improving resolution where the bound-state wavefunctions are localized.
We use $\gamma = 26.8$ to resolve the wavefunctions sufficiently.
Increasing $N$ generally improves accuracy.

\begin{figure}
    \centering
    \includegraphics[width=1.0\linewidth]{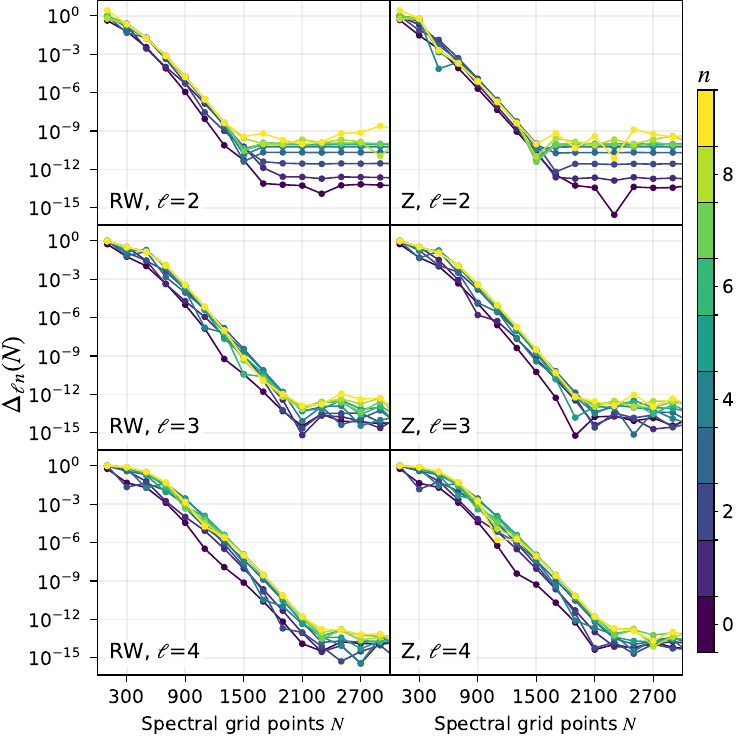}
    \caption{Relative differences between spectral and shooting-method eigenvalues for the inverted Regge-Wheeler (left) and Zerilli (right) potentials. Rows show $\ell=2,3,4$ from top to bottom; color indicates energy level $n$. We plot $\Delta_{\ell n}(N)=|E^{\mathrm{spectral}}_{\ell n}(N)-E^{\mathrm{shooting}}_{\ell n}|/|E^{\mathrm{shooting}}_{\ell n}|$, with shooting energies computed at $\mathrm{rtol}=10^{-17}$.}
    \label{fig:summary_spectral_grid_convergence}
\end{figure}

The boundary location determines how many bound states can be resolved accurately.
A smaller domain captures low-lying states well but may miss shallow states whose wavefunctions extend beyond the boundaries.
This is illustrated in Figs.~\ref{summary_spectral_self_convergence_l234_two_column} and~\ref{fig:summary_spectral_grid_convergence}.
For $\ell=2$, higher states become progressively less accurate within our domain, remaining accurate to $\sim 10^{-8}$ or better.
For $\ell=3,4$, the deeper potential wells allow more states to be resolved to higher accuracy within the same domain.

We verify convergence by checking that eigenvalues stabilize as $N$ increases and that pseudospectral and shooting results agree within the prescribed tolerance.
Figure~\ref{summary_spectral_self_convergence_l234_two_column} shows that convergence plateaus occur at different grid resolutions for different $n$, typically below $10^{-12}$ for small $n$, especially for $\ell=2$, while large-$n$ eigenvalues remain poorly resolved.

Figure~\ref{fig:summary_spectral_grid_convergence} compares shooting-method eigenvalues (at $\mathrm{rtol}=10^{-17}$) with pseudospectral eigenvalues as $N$ increases from $100$ to $3100$.
The two methods agree closely for low-lying states but differ for higher states due to the finite computational boundaries; resolving these would require larger domains and correspondingly more grid points.

\begin{figure}
    \centering
    \includegraphics[width=1.0\linewidth]{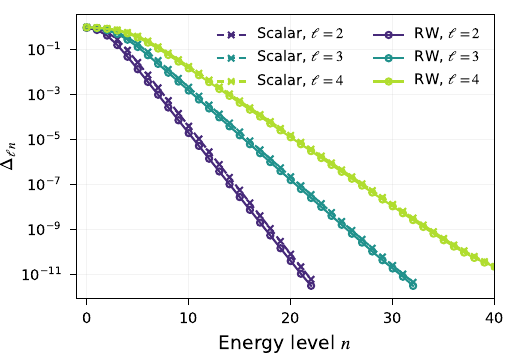}
    \caption{Relative deviations of the shooting-method eigenvalues from Hod's asymptotic analytical expression~\cite{Hod:2025scz} for the inverted Regge-Wheeler and Scalar potentials. The plotted quantity is
    $\Delta_{\ell n}=|E_{\ell n}^{\mathrm{numerical}}-E_{\ell n}^{\mathrm{analytic}}|/|E_{\ell n}^{\mathrm{analytical}}|$.
    Results are shown for $\ell=2,3,4$.}
    \label{fig:summary_rw_scalar_hod_analytical_error_l234_single_column}
\end{figure}

Figure~\ref{fig:summary_rw_scalar_hod_analytical_error_l234_single_column} shows the relative deviation of the shooting energies (at $\mathrm{rtol}=10^{-17}$) from the asymptotic spectrum of Ref.~\cite{Hod:2025scz}, valid for $n\gg\ell$, for both the inverted Regge-Wheeler and scalar potentials.
The deviation generally decreases with increasing $n$, as expected from the improving asymptotic approximation.
Since the reference spectrum is itself asymptotic, this comparison mixes numerical and analytical errors and is not a direct accuracy measure.
It is nonetheless complementary to the pseudospectral check: the pseudospectral method is most reliable at low $n$, while Hod's expression is most reliable at high $n$, together validating the shooting-method results across the full spectrum.

\section{Additional Results}\label{app_add_results}

Table~\ref{tab:master-bound-state-data} lists the numerically computed bound-state energies of the inverted Regge-Wheeler, Zerilli, and scalar-field potentials for selected $\ell$ and $n$.
At fixed $\ell$, the relative difference between the Regge-Wheeler and Zerilli energies appears to approach a constant as $n$ increases.
This asymptotic value decreases by roughly an order of magnitude with each unit increase in $\ell$.

\begin{figure}[H]
    \centering
    \includegraphics[width=1.0\linewidth]{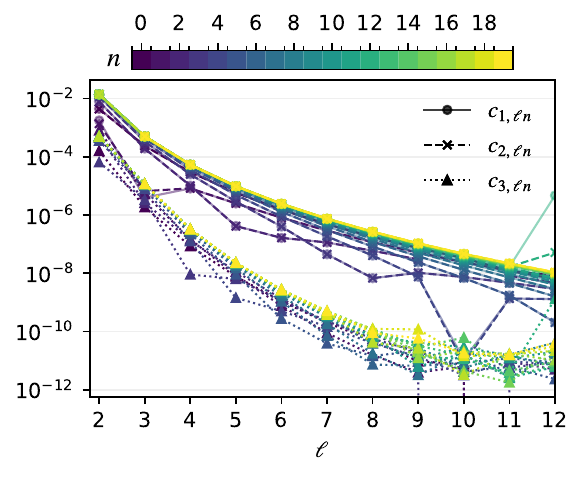}
    \caption{Third order taylor polynomial  to parabolic relative energy shift $\Delta E = |E_{\ell n}(\epsilon) -E_{\ell n}(0)|/|E_{\ell n}(0)|$ in interpolated potential . $c_{k,\ell n} = d_k/(k!\, E_{\ell n}(0))$. First order coefficients (solid lines) and second order coefficients (dashed lines) overlap while third order coefficients (dotted lines) are much smaller, implying $|c_1| \approx |c_2| \gg |c_3|$. }
    \label{fig:summary_parabola_fit_coefficients}
\end{figure}

Figure~\ref{fig:summary_parabola_fit_coefficients} plots the Taylor coefficients of the parabolic energy shift for different $ell$ and $n$. These coefficients serve as an excellent proxy for understanding the structure of the spectral splitting. An interpretation of this is provided in Sec.~\ref{app_spectral_splitting}.

\begin{figure}[H]
    \centering
    \includegraphics[width=1.0\linewidth]{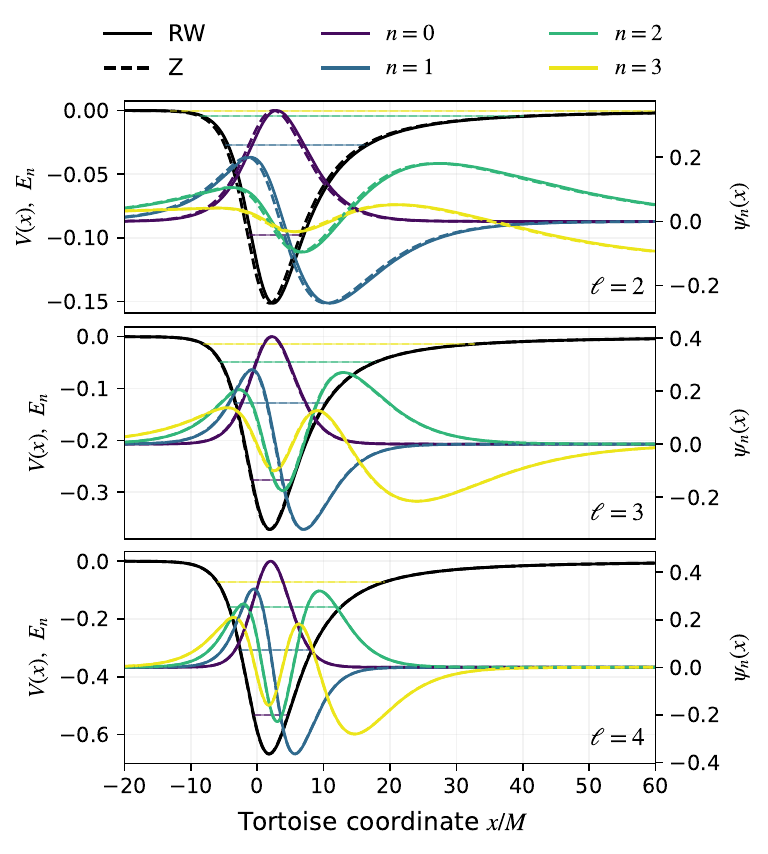}
    \caption{First four wavefunction for $\ell= 2, \,3,\, 4$, for inverted Regge-Wheeler (solid lines) and Zerilli (dashed lines) potentials.}
    \label{fig:summary_eigenfunctions}
\end{figure}

Figure~\ref{fig:summary_eigenfunctions} shows the Regge-Wheeler and Zerilli eigenfunctions for selected $\ell$ and $n$.
At fixed $n$, the eigenfunctions become increasingly similar as $\ell$ increases, while increasing $n$ leads to rapid delocalization.
The deeper wells at larger $\ell$ also yield more strongly bound states at fixed $n$.
The small relative displacement between the potentials is reflected in their eigenfunctions.
Although both potentials share the same large-$r$ asymptotic form, their differences cannot be eliminated by a simple shift.

\begin{table*}
\caption{Bound-state energies, Zerilli-to-Regge comparisons, and phase offsets for $M=1$.}
\label{tab:master-bound-state-data}
\begin{tabular}{cc|c|c|c|c|c|c|c}
\toprule
$\ell$ & $n$ & $E^{\rm RW}_n$ & $E^{\rm Z}_n$ & $E^{\rm Scalar}_n$ & $|E_n^Z/E_n^{RW}-1|$ & $\delta^{\rm RW}_n$ & $\delta^{\rm Z}_n$ & $\delta^{\rm Scalar}_n$ \\
\hline
2 & 0 & -9.74078306e-02 & -9.74141976e-02 & -1.69393082e-01 & 6.53650230e-05 & 0.8887821499 & 0.8887572048 & 0.6776148029 \\
 & 1 & -2.67681591e-02 & -2.67580181e-02 & -5.77188106e-02 & 3.78843724e-04 & 0.3817442526 & 0.3818888619 & 0.0885033980 \\
 & 2 & -4.47439568e-03 & -4.47403435e-03 & -1.27243173e-02 & 8.07565406e-05 & 0.0644380947 & 0.0644689158 & -0.3344306659 \\
 & 3 & -4.72771021e-04 & -4.72894176e-04 & -1.68697780e-03 & 2.60497388e-04 & -0.0778197958 & -0.0779191991 & -0.5632976435 \\
 & 4 & -3.89683675e-05 & -3.89843376e-05 & -1.54819847e-04 & 4.09820532e-04 & -0.1252990595 & -0.1254554315 & -0.6517761978 \\
 & 5 & -2.93973435e-06 & -2.94107616e-06 & -1.21009724e-05 & 4.56440965e-04 & -0.1389764304 & -0.1391505868 & -0.6789844872 \\
 & 6 & -2.16078759e-07 & -2.16180216e-07 & -8.98518078e-07 & 4.69534787e-04 & -0.1427309587 & -0.1429101099 & -0.6866078039 \\
 & 7 & -1.57688936e-08 & -1.57763539e-08 & -6.57539854e-08 & 4.73105594e-04 & -0.1437482872 & -0.1439288006 & -0.6886850188 \\
 & 8 & -1.14853792e-09 & -1.14908241e-09 & -4.79283938e-09 & 4.74071712e-04 & -0.1440230396 & -0.1442039215 & -0.6892468428 \\
 & 9 & -8.36105665e-11 & -8.36502257e-11 & -3.48977096e-10 & 4.74332554e-04 & -0.1440971825 & -0.1442781640 & -0.6893985110 \\
 & 10 & -6.08576791e-12 & -6.08865501e-12 & -2.54024117e-11 & 4.74402940e-04 & -0.1441171863 & -0.1442981946 & -0.6894394352 \\
 & 16 & -9.04716538e-19 & -9.05145762e-19 & -3.77642521e-18 & 4.74428936e-04 & -0.1441245741 & -0.1443055923 & -0.6894545498 \\
 & 17 & -6.58482121e-20 & -6.58794524e-20 & -2.74860510e-19 & 4.74428943e-04 & -0.1441245762 & -0.1443055944 & -0.6894545540 \\
 & 18 & -4.79264702e-21 & -4.79492079e-21 & -2.00052418e-20 & 4.74428945e-04 & -0.1441245768 & -0.1443055950 & -0.6894545552 \\
 & 19 & -3.48824436e-22 & -3.48989929e-22 & -1.45604656e-21 & 4.74428946e-04 & -0.1441245769 & -0.1443055951 & -0.6894545555 \\
 & 20 & -2.53885768e-23 & -2.54006219e-23 & -1.05975804e-22 & 4.74428946e-04 & -0.1441245770 & -0.1443055952 & -0.6894545556 \\
 & $\infty$ & - & - & - & - & -0.1441245770 & - & -0.6894545556 \\
\hline
3 & 0 & -2.76400454e-01 & -2.76400696e-01 & -3.56404229e-01 & 8.76094576e-07 & 0.7015325017 & 0.7015320238 & 0.5628441633 \\
 & 1 & -1.28348117e-01 & -1.28347785e-01 & -1.76141030e-01 & 2.58776678e-06 & 0.1200307003 & 0.1200321121 & -0.0526588034 \\
 & 2 & -4.84009788e-02 & -4.84009262e-02 & -7.20990448e-02 & 1.08588065e-06 & -0.3479291270 & -0.3479285346 & -0.5653443825 \\
 & 3 & -1.42072395e-02 & -1.42072647e-02 & -2.32570800e-02 & 1.77445621e-06 & -0.6792042513 & -0.6792052194 & -0.9480856913 \\
 & 4 & -3.26581293e-03 & -3.26582543e-03 & -5.82043988e-03 & 3.82650489e-06 & -0.8771048940 & -0.8771069815 & -1.1923635736 \\
 & 5 & -6.27250173e-04 & -6.27253199e-04 & -1.18332561e-03 & 4.82468489e-06 & -0.9769825014 & -0.9769851336 & -1.3232677409 \\
 & 6 & -1.08883081e-04 & -1.08883651e-04 & -2.11608266e-04 & 5.23626861e-06 & -1.0216759156 & -1.0216787723 & -1.3841769776 \\
 & 7 & -1.80229887e-05 & -1.80230860e-05 & -3.54992641e-05 & 5.39911115e-06 & -1.0404250141 & -1.0404279596 & -1.4102374874 \\
 & 8 & -2.92315160e-06 & -2.92316757e-06 & -5.79014303e-06 & 5.46350682e-06 & -1.0480670490 & -1.0480700296 & -1.4209514202 \\
 & 9 & -4.70157274e-07 & -4.70159855e-07 & -9.33422670e-07 & 5.48909994e-06 & -1.0511457173 & -1.0511487119 & -1.4252828003 \\
 & 10 & -7.53645363e-08 & -7.53649507e-08 & -1.49762838e-07 & 5.49930522e-06 & -1.0523803655 & -1.0523833657 & -1.4270222430 \\
 & 26 & -1.38318145e-20 & -1.38318906e-20 & -2.75033071e-20 & 5.50609582e-06 & -1.0532041990 & -1.0532072029 & -1.4281836549 \\
 & 27 & -2.21217854e-21 & -2.21219072e-21 & -4.39871615e-21 & 5.50609581e-06 & -1.0532041992 & -1.0532072031 & -1.4281836552 \\
 & 28 & -3.53802741e-22 & -3.53804689e-22 & -7.03504625e-22 & 5.50609582e-06 & -1.0532041993 & -1.0532072032 & -1.4281836553 \\
 & 29 & -5.65851162e-23 & -5.65854278e-23 & -1.12514366e-22 & 5.50609581e-06 & -1.0532041993 & -1.0532072032 & -1.4281836554 \\
 & 30 & -9.04988856e-24 & -9.04993839e-24 & -1.79948817e-23 & 5.50609582e-06 & -1.0532041993 & -1.0532072032 & -1.4281836554 \\
 & $\infty$ & - & - & - & - & -1.0532041994 & - & -1.4281836554 \\
\hline
4 & 0 & -5.32510132e-01 & -5.32510155e-01 & -6.16755063e-01 & 4.16453829e-08 & 0.4457074912 & 0.4457074618 & 0.3418263071 \\
 & 1 & -3.07505500e-01 & -3.07505477e-01 & -3.65653574e-01 & 7.27425289e-08 & -0.1659077254 & -0.1659076740 & -0.2884074199 \\
 & 2 & -1.59071488e-01 & -1.59071478e-01 & -1.95541021e-01 & 6.51825517e-08 & -0.6996985004 & -0.6996984543 & -0.8456968804 \\
 & 3 & -7.19259371e-02 & -7.19259369e-02 & -9.20156581e-02 & 3.10085690e-09 & -1.1383025971 & -1.1383025949 & -1.3125260302 \\
 & 4 & -2.79471379e-02 & -2.79471396e-02 & -3.73382314e-02 & 6.31298391e-08 & -1.4696762064 & -1.4696762511 & -1.6745831357 \\
 & 5 & -9.35362965e-03 & -9.35363068e-03 & -1.30125560e-02 & 1.09726597e-07 & -1.6955007190 & -1.6955007966 & -1.9290160195 \\
 & 6 & -2.76688274e-03 & -2.76688311e-03 & -3.97205019e-03 & 1.35523685e-07 & -1.8339787556 & -1.8339788515 & -2.0897109080 \\
 & 7 & -7.51186820e-04 & -7.51186932e-04 & -1.10083001e-03 & 1.48088862e-07 & -1.9117852517 & -1.9117853565 & -2.1820908176 \\
 & 8 & -1.93653081e-04 & -1.93653111e-04 & -2.87197824e-04 & 1.53955965e-07 & -1.9529791738 & -1.9529792827 & -2.2317282022 \\
 & 9 & -4.85223008e-05 & -4.85223084e-05 & -7.24236345e-05 & 1.56705768e-07 & -1.9740442254 & -1.9740443363 & -2.2573243878 \\
 & 10 & -1.19789157e-05 & -1.19789175e-05 & -1.79385748e-05 & 1.58015926e-07 & -1.9846195819 & -1.9846196937 & -2.2702311126 \\
 & 36 & -1.31933948e-21 & -1.31933969e-21 & -1.98224995e-21 & 1.59256166e-07 & -1.9950259075 & -1.9950260201 & -2.2829686841 \\
 & 37 & -3.20877473e-22 & -3.20877524e-22 & -4.82104388e-22 & 1.59256166e-07 & -1.9950259075 & -1.9950260202 & -2.2829686842 \\
 & 38 & -7.80408332e-23 & -7.80408456e-23 & -1.17252943e-22 & 1.59256166e-07 & -1.9950259075 & -1.9950260202 & -2.2829686842 \\
 & 39 & -1.89803653e-23 & -1.89803683e-23 & -2.85171698e-23 & 1.59256163e-07 & -1.9950259076 & -1.9950260202 & -2.2829686843 \\
 & 40 & -4.61622783e-24 & -4.61622857e-24 & -6.93568069e-24 & 1.59256160e-07 & -1.9950259076 & -1.9950260202 & -2.2829686843 \\
 & $\infty$ & - & - & - & - & -1.9950259076 & - & -2.2829686843 \\
\hline
\end{tabular}
\end{table*}

\end{document}